\documentclass[manuscript]{acmart}

\AtBeginDocument{%
  }

\copyrightyear{2025}
\acmYear{2025}
\setcopyright{rightsretained}
\acmConference[CHI EA '25]{Extended Abstracts of the CHI Conference on Human Factors in Computing Systems}{April 26-May 1, 2025}{Yokohama, Japan}
\acmBooktitle{Extended Abstracts of the CHI Conference on Human Factors in Computing Systems (CHI EA '25), April 26-May 1, 2025, Yokohama, Japan}\acmDOI{10.1145/3706599.3719760}
\acmISBN{979-8-4007-1395-8/2025/04}
\acmISBN{978-1-4503-XXXX-X/18/06}

\begin{document}

\title{CyanKitten: AI-Driven Markerless Motion Capture for Improved Elderly Well-Being}

\author{Mengyao Guo}
 \authornote{Both authors contributed equally to this research.}
\affiliation{%
 \institution{Harbin Institute of Technology (Shenzhen)}
  \city{Shenzhen}
  \country{China}}
\email{guomengyao@hit.edu.cn}

\author{Yu Nie}
\authornotemark[1]
\affiliation{%
  \institution{Beijing Normal University \& Cyanpuppets}
  \city{Zhuhai\&Guangzhou}
  \country{China}}
\email{nieyu@cyanpuppets.com}

\author{Jinda Han}
\affiliation{%
  \institution{University of Illinois at Urbana-Champaign}
  \city{Urbana}
  \country{United States}}
\email{jhan51@illinois.edu}

\author{Zongxing Li}
\affiliation{%
  \institution{Cyanpuppets}
  \city{Guangzhou}
  \country{China}}
\email{epik@cyanpuppets.com}

\author{Ze Gao}
\authornote{Ze Gao is the corresponding author.}
\affiliation{%
  \institution{Cyanpuppets \& Hong Kong Polytechnic University}
  \city{Guangzhou \& Hong Kong SAR }
  \country{China}}
\email{zegaoap@cyanpuppets.com  \&  zegaoap@hotmail.com}

\renewcommand{\shortauthors}{Guo et al.}

\begin{abstract}

This paper introduces CyanKitten, an interactive virtual companion system tailored for elderly users, integrating advanced posture recognition, behavior recognition, and multimodal interaction capabilities. The system utilizes a three-tier architecture to process and interpret user movements and gestures, leveraging a dual-camera setup and a convolutional neural network specifically trained on elderly movement patterns. The behavior recognition module identifies and responds to three key interactive gestures: greeting waves, petting motions, and heart-making gestures. Additionally, a multimodal integration layer combines visual and audio inputs to facilitate natural and intuitive interactions. This paper outlines the technical implementation of each component, addressing challenges such as elderly-specific movement characteristics, real-time processing demands, and environmental adaptability. The result is an engaging and accessible virtual interaction experience designed to enhance the quality of life for elderly users.

\end{abstract}



\begin{teaserfigure}
  \includegraphics[width=\textwidth]{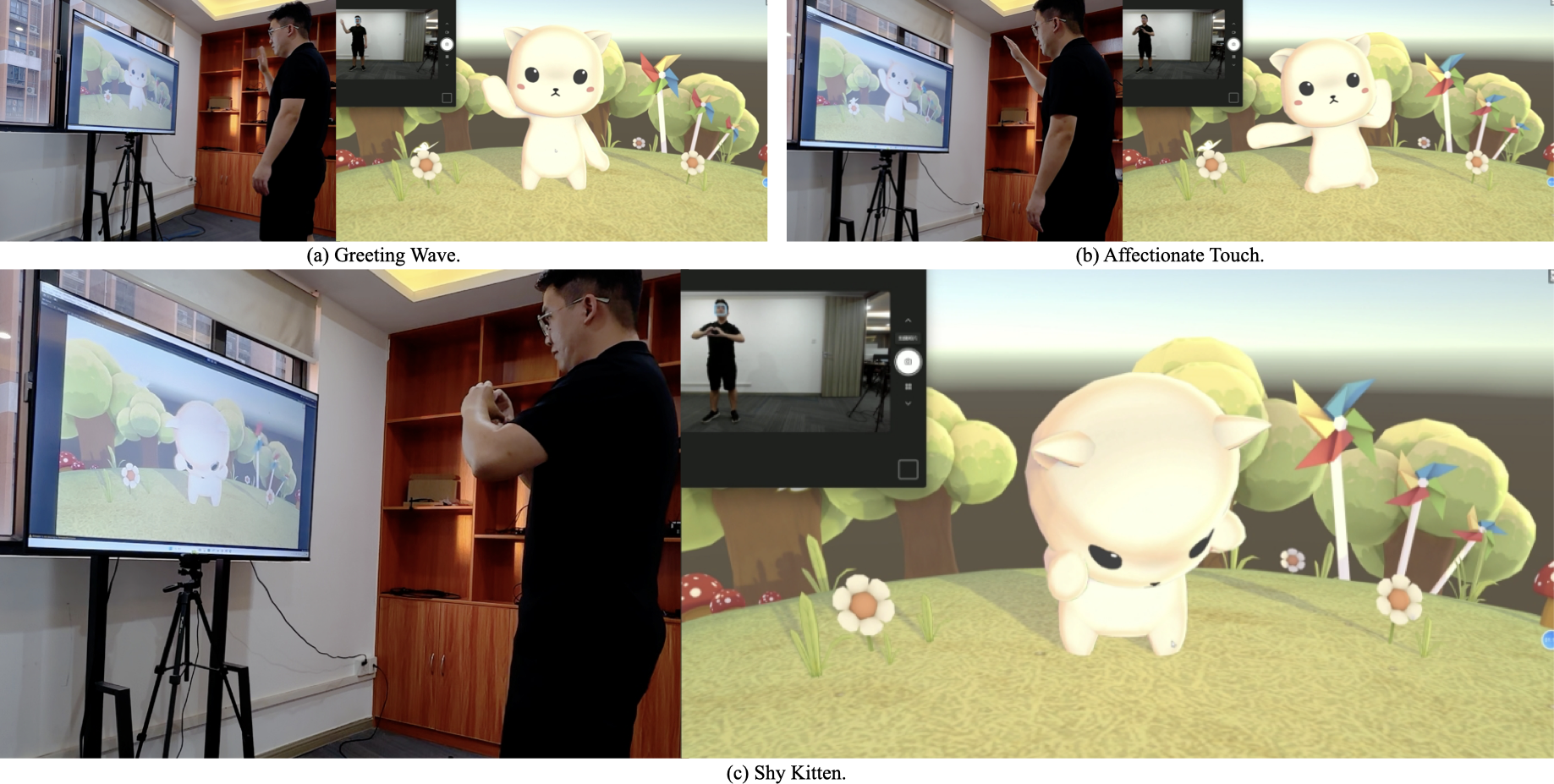}
  \caption{The Corresponding Gestures of CyanKitten.}
  \Description{The Corresponding Gestures.
  \copyright Credit by Authors.}
  \label{fig:teaser}
\end{teaserfigure}

\maketitle

\section{Introduction}

China has witnessed a significant demographic shift in recent years characterized by an increasing life expectancy~\cite{fang2020research}. While indicative of improved health and social conditions, this change presents substantial societal challenges, particularly in elderly care~\cite{peng2023negative}. According to the National Bureau of Statistics of China~\footnote{National Bureau of Statistics of China: https://www.stats.gov.cn/english/}, by 2033, the elderly population in China is projected to surpass 400 million, accounting for about one-fourth of the total population. This figure is expected to peak at 487 million by 2053, representing more than one-third of the population~\footnote{China Research Center on Aging: http://www.crca.cn/index.php/13-agednews/842-2023-03-10-04-12-47.html}. Such growth strains an already burdened healthcare system and necessitates innovative solutions to manage the escalating demand for elderly care services.

However, the current technological offerings for the elderly predominantly focus on healthcare monitoring and management, neglecting broader applications that could enhance daily living and emotional well-being~\cite{blackman2016ambient}. At the same time, due to the consequence of both societal neglect and the diminishing physical presence of their children, many elderly individuals feel increasingly isolated. On the one hand, there is a palpable sense of abandonment and loneliness, feelings that are exacerbated by the minimal interpersonal interactions~\cite{pound2016human}. On the other hand, the elderly are curious about technology, particularly advancements in artificial intelligence (AI)~\cite{shandilya2022understanding}. So, we consider integrating AI into daily life through interactive applications such as digital companions, which could alleviate feelings of isolation and provide meaningful engagement.

By considering both the technological and emotional needs of the elderly, we designed our project, CyanKitten, which aims to create a more inclusive and supportive environment for the elderly society, offering them a new way to embrace technology and its benefits. 

\section{Related Works}
At this stage there are actually many examples of using motion capture to create visual graphics\cite{gao2022meditation,gao2022metanalysis}. To apply our markerless motion capture technology to elderly care, we analyzed the related works that inspired us to integrate digital avatars to interact with old people to foster an active and enriching lifestyle and indirectly bolster health by alleviating stress, depression, and anxiety, conditions prevalent in isolated seniors.

\begin{figure}
\centering
\includegraphics[width=\linewidth]{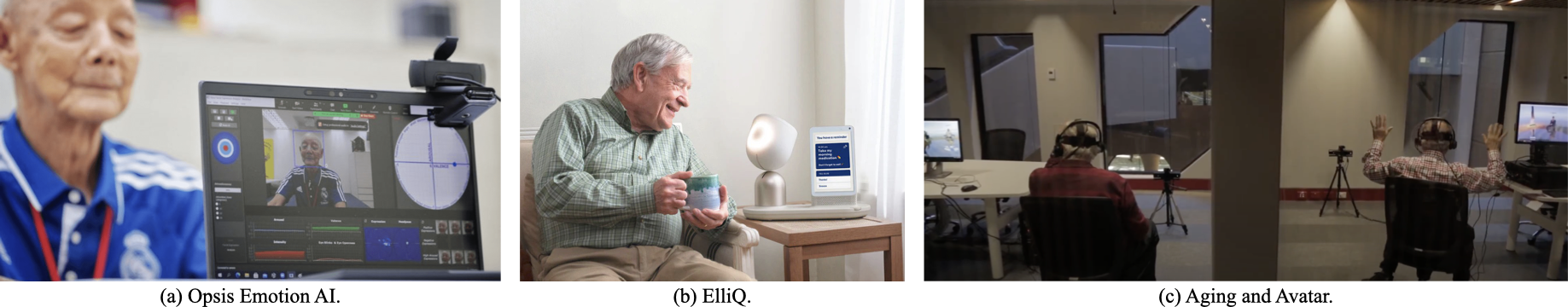}
\caption{The Related Works.
\copyright Credit by Mentioned Companies.}
\Description{The Related Works.}
\label{related}
\end{figure}

Opsis Emotion AI~\footnote{Opsis Emotion AI: https://opsis.ai/} (Figure~\ref{related}a) offers solutions for analyzing and interpreting human emotions using cutting-edge computer vision and signal processing technologies, including the circumplex emotion detection algorithm, the large language model, and multimodal sensor technology, allowing for a comprehensive understanding of complex human emotional states. Opsis Emotion AI can analyze the emotional states of seniors through video calls~\cite{gunasegaran2021ai} by capturing facial expressions and responses, which are then utilized by counselors at Lions Befrienders to diagnose mental health conditions such as anxiety and depression more accurately. Although this method facilitates early detection and intervention, enhancing the quality of virtual mental health care provided to elderly patients, it is limited to the elderly who actively communicate and have psychological needs and fails to reach more potential users who often need more attention at the initial stage. and it focuses on diagnostic support rather than directly treating the emotional and psychological needs of the elderly.

ElliQ~\footnote{ElliQ: https://elliq.com/} is an AI-powered companion robot (Figure~\ref{related}b) designed specifically for the elderly, providing company and proactive support to improve their daily lives. The robot combines voice interaction and physical movement to engage users, such as delivering reminders of medication, facilitating communication with family, and encouraging physical activity~\cite{becchimanzi20227}. It helps reduce feelings of loneliness and isolation, but its persistent reminder function might inadvertently emphasize to elderly users their dependency and old age, potentially affecting their sense of autonomy. Additionally, ElliQ required costly payments followed by monthly subscriptions.

The~\textit{Ageing and Avatars}~\footnote{Ageing and Avatars: https://cis.unimelb.edu.au/hci/projects/ageing-avatars} project (Figure~\ref{related}c) at The University of Melbourne is dedicated to enhancing social participation among the elderly through the creation of a virtual presence using virtual and augmented reality technologies (VR/AR). They consider maintaining social networks and active community participation for older individuals, as these factors contribute to enhanced well-being and a more fulfilling life in advanced age~\cite{baker2020evaluating} for the reason that with increasing age, opportunities for such engagement often diminish due to growing frailty and health-related mobility restrictions, confining many olds to homes. So they focus on how these adults use virtual avatars to represent themselves, aiming to foster continued social interaction and reduce isolation among the aging population based on the full-bodied gesture-based interactions and avatars, allowing older adults to establish a sense of presence and interact in social virtual environments.

In conclusion, avatars empower users to transcend real-world physical limitations, facilitating engagement in social environments via digital platforms that require less physical effort and offer more control. The elderly often confront overlooked challenges of aging, including emotional health and social connectivity. These technologies deliver continuous engagement and a sense of purpose by prioritizing companionship and proactive support over direct medical treatment. This strategy not only fosters an active and enriching lifestyle but also indirectly bolsters health by alleviating stress, depression, and anxiety, conditions prevalent in isolated seniors.

\section{Design of CyanKitten}
The technology~\cite{guo2024technical} serves as our foundation for the CyanKitten project, a groundbreaking initiative designed to enhance the quality of life for the elderly. This project integrates our existing markerless motion capture technology in an entertainment way, creating interactive digital avatars tailored for the elderly demographic. CyanKitten aims to address critical issues such as loneliness, social isolation, and cognitive decline among older adults. We have developed three interactive gestures that correspond between human and cat motions. Moving forward, we will thoroughly consider and discuss the ethical issues, privacy concerns, and psychological impacts of replacing human interaction with avatars by applying for IRB approval and conducting semi-structured user studies and interviews.

\begin{figure}
\centering
\includegraphics[width=\linewidth]{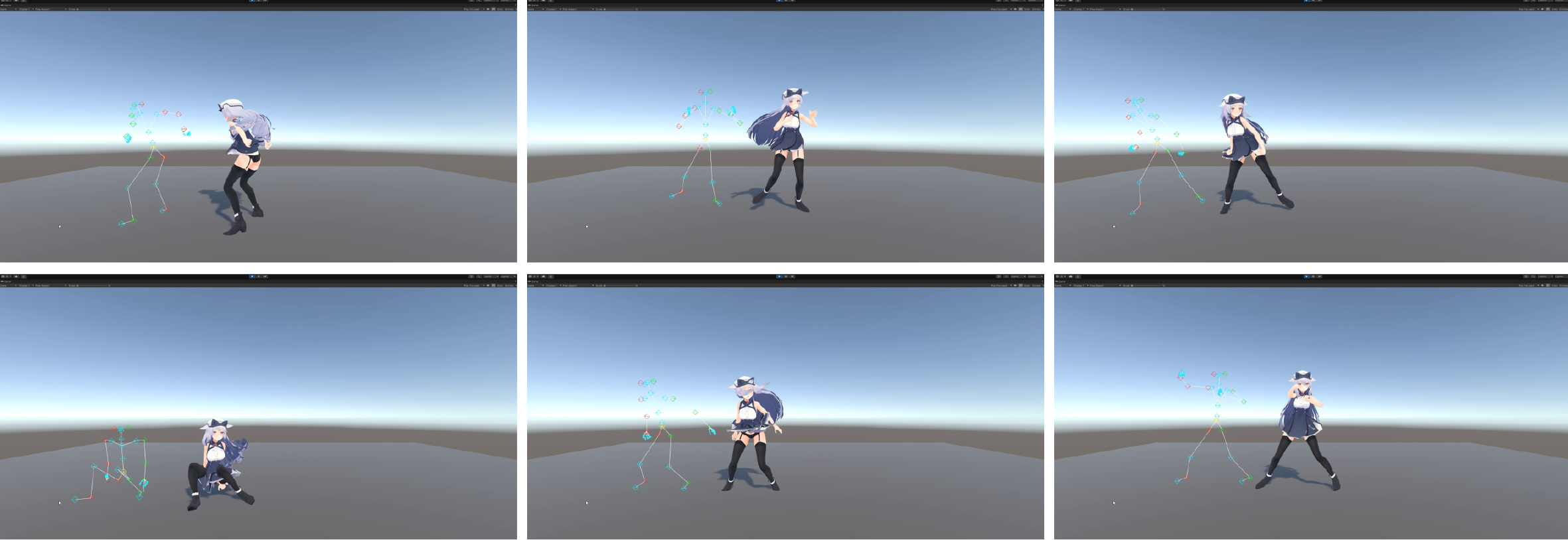}
\caption{The Display of Real-Time Motion Capture.
\copyright Credit by Authors.}
\Description{The Display of Real-Time Motion Capture.}
\label{motion}
\end{figure}

\section{Technical Implementation}
\begin{figure}
\centering
\includegraphics[width=\linewidth]{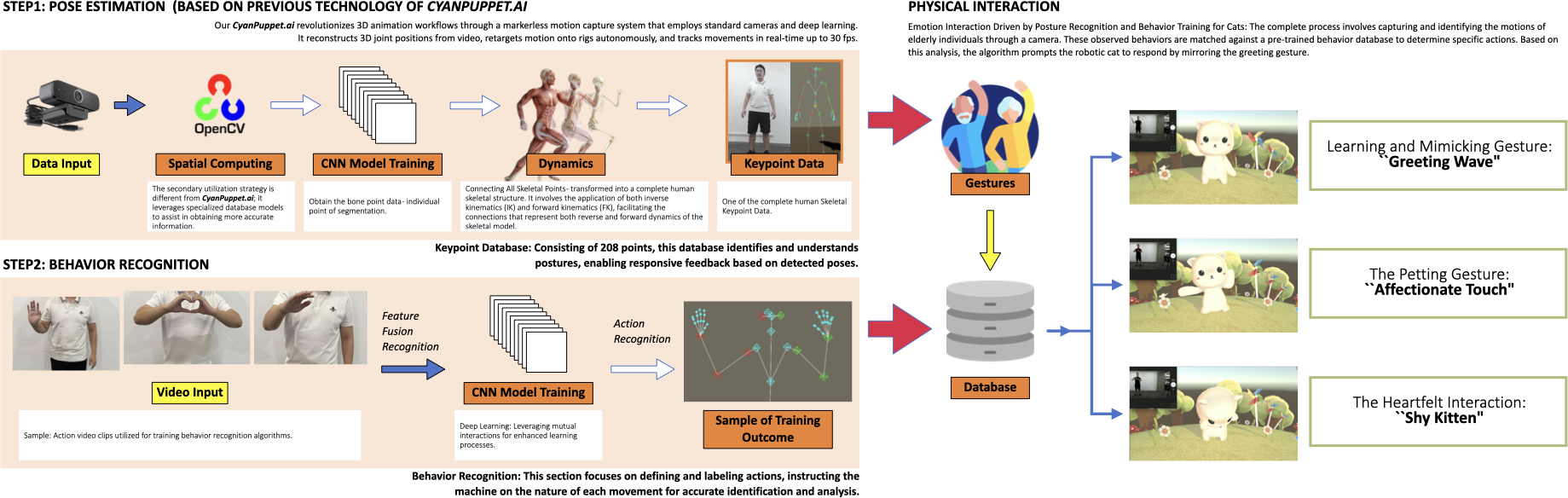}
\caption{The Display of Real-Time Motion Capture.
\copyright Credit by Authors.}
\Description{The Display of Real-Time Motion Capture.}
\label{workflow}
\end{figure}

We utilize NVIDIA CUDA GPU technology to facilitate the training of extensive datasets of human poses over extended periods, a critical aspect of handling the complexities of human motion. Our system proficiently captures both the spatial and temporal dependencies characteristic of human movements by integrating state-of-the-art convolutional neural networks (CNNs) and recurrent neural networks (RNNs). During real-time operations, these deep-learning models can extract up to 208 essential 3D joint locations from streamed stereo video frames. The key points detected are systematically organized into a skeletal structure that strictly adheres to human biomechanical constraints, ensuring the captured motion is both accurate and realistic (Figure~\ref{workflow}).


\subsection{Motion Capture}

Through the enhancement of~\textbf{Spatial Computing} and~\textbf{Automatic Motion Retargeting}, we have developed~\textbf{Specialized Algorithms}, including the Cyan\_SLAM algorithm for initial model matching, dynamic real-time adjustments of model proportions, and an algorithm for self-adaptive conversion based on original bone topologies. These advancements streamline the process, reduce the need for manual mapping, and enhance the efficiency of producing accurate and realistic 3D animations.

Our approach incorporates advanced spatial computing techniques utilizing a dual-camera setup based on stereo vision principles. This setup employs feature matching and stereo correspondence algorithms, enabling depth reconstruction from the two-dimensional image planes captured by the cameras. The process involves thoroughly calibrating each camera to accurately model its lens properties and determine their spatial relationship. This calibration facilitates the triangulation of corresponding 2D image points into precise 3D world coordinates. Further leveraging deep learning tools like OpenPose~\cite{8765346} and PoseNet~\cite{kendall2015posenet}, which utilize recent advances in neural networks for accurate 2D human pose estimation, our system can identify and track 32 key anatomical landmarks on the human body, providing robust tracking of joint orientations and positions over time~\cite{guo2024technical}.

Addressing the challenge of maintaining consistency in motion capture data, our project employs a machine learning strategy to automate the retargeting process. Traditional methods often involve labor-intensive adjustments to reconcile differences in bone structures, names, and hierarchies across various 3D models. In contrast, our system uses a deep learning framework to analyze and learn from a vast array of motion data, allowing it to adapt and provide dynamic feedback. 

\subsection{Motion Interaction}

In digital companionship and therapeutic interaction, our project explores the unique dynamics between elderly individuals and digital pets, capitalizing on cats' natural behaviors because the avatars reduce loneliness and cognitive decline among the elderly~\cite{cyarto2016active}. Cats seldom communicate with each other through vocalizations; instead, they rely predominantly on body movements. This characteristic makes cats well-suited for our motion capture system, which focuses on motion rather than sound for communication. Unlike dogs, which are often anthropomorphized through training rather than inherent animal behavior, cats offer a more authentic and spontaneous interaction model. Therefore, we chose a cat avatar from the AssetStore called Pspsps Cat as ``CyanKitten," to communicate through gestures and movements, creating a non-verbal interaction system that is intuitive and comforting for the elderly.

The interaction design of our system is tailored to invoke a sense of being needed and an emotional bond through mutual care and critical psychological support for elderly individuals who might feel isolated or less required by their families. For instance, when an elderly person performs a specific gesture, like raising a hand, the cat avatar responds by mimicking this gesture or performing a complementary action, such as stretching up or playfully reaching out, which simulates the nurturing behavior typically seen in parent-child dynamics. This interactive learning and mimicking process not only engages the elderly but also helps rekindle their feelings of being caregivers and educators, roles that they might miss in their current stage of life.

Furthermore, we incorporate cat-like motions corresponding to human gestures to enhance the interaction. The elderly make a gentle motion of petting, which can lead the avatar to deliver a joyful expression. This reaction fosters an emotional connection, making the elderly feel deeply understood and valued. This corresponding reaction strengthens the uniqueness of the interaction, making them feel special in their ability to communicate in a ``cat's language." Such interactions are designed not just for entertainment but also to potentially incorporate health benefits, like encouraging physical gestures that can benefit the body health of the elderly.

\subsection{Reaction to Human Gesture}

We present three initial specific interactions in our project that foster a meaningful connection between elderly individuals and the CyanKitten using our motion capture system (see Figure~\ref{fig:teaser}):




\begin{itemize}

    \item \textbf{Learning and Mimicking Gesture:}
\textit{``Greeting Wave"}

\textbf{Human Gesture}: The elderly user raises their hand in a ``say hi" gesture.
\textbf{Cat Response}: The avatar mimics this gesture by stretching its front paws upward or performing the same motion at users. This interaction is designed to replicate the dynamic of an elder teaching a younger individual, giving the elderly a sense of purpose and rekindling their role as mentors.

    \item \textbf{The Petting Gesture:}
\textit{``Affectionate Touch"}

\textbf{Human Gesture}: The elderly user makes a gentle motion in the air, mimicking the act of petting and stroking.
\textbf{Cat Response}: The avatar displays a joyful ``victory" expression, waves its hands, and smiles. This reaction fosters a profound emotional connection, making the elderly feel deeply understood and valued. 

    \item \textbf{The Heartfelt Interaction:}
\textit{``Shy Kitten"}

\textbf{Human Gesture}: The elderly user forms a heart shape with her/his hands.
\textbf{Cat Response}: In response, the avatar displays a shy expression. This reaction captures the gesture's affectionate intent. It enhances the interactive experience, making it more engaging and emotionally resonant for the user and encouraging enjoyable physical activity.

\end{itemize}

CyanKitten utilizes webcams to capture human gestures, which are then analyzed using multi-layer neural networks to achieve precise posture and behavior recognition. The system allows machines to interpret human behaviors by accurately identifying human postures. The interactions designed within this framework are meant to replicate the typical dynamics between humans and cats, thereby enhancing the emotional and physical well-being of elderly users. Engaging in these interactions can provide the elderly with joy, companionship, and a renewed sense of purpose.

\section{User Study Plan}

We are going to implement a comprehensive evaluation of CyanKitten in community centers for older adults using fuzzy-set Qualitative Comparative Analysis (fsQCA) combined with traditional assessment methods. The study is going to involve at least 30 elderly participants (aged 65+) recruited from three community centers in urban China (our city), where we are going to set up dedicated interaction spaces equipped with our motion capture system and display screens. Each center will host at least 10 participants for a 12-week study period to guarantee that we collected enough data, allowing for controlled environment testing while maintaining a natural community setting.

The study is going to examine multiple causal conditions that potentially contribute to successful engagement with CyanKitten's motion-based interactions. These conditions are going to include participants' physical mobility levels (assessed through standardized tests), technology acceptance (measured via questionnaires), social engagement levels (evaluated through activity participation records), cognitive function scores (using MoCA), and demographic factors. Our outcome measures are going to focus on interaction quality (gesture recognition accuracy and completion rates), engagement levels (frequency and duration of voluntary interactions), and emotional well-being improvements (assessed through standardized scales and behavioral observations). Data collection is going to combine quantitative metrics from the motion capture system (interaction frequency, gesture accuracy, response time) with qualitative assessments (semi-structured interviews, observation notes) conducted during thrice-weekly sessions. Each participant is going to attend scheduled 30-minute sessions, where they can freely interact with CyanKitten through our three core gestures: greeting wave, affectionate touch, and heartfelt interaction. The community center setting will allow for both structured evaluation periods and spontaneous interactions, providing rich data on natural engagement patterns.

The fsQCA methodology is going to help identify which combinations of conditions lead to high engagement and positive emotional outcomes. This approach will be particularly suitable for understanding the complex pathways to successful technology adoption among elderly users, as it can reveal multiple sufficient configurations rather than assuming a single optimal solution. We are going to calibrate condition and outcome measures into fuzzy-set membership scores (0.0-1.0) based on theoretical and substantive knowledge of elderly care and technology interaction. Our analysis is going to proceed through systematic case comparison and boolean minimization to identify necessary and sufficient conditions for successful engagement. This will be complemented by traditional statistical analysis of motion capture data and thematic analysis of qualitative feedback. The study will operate under strict ethical guidelines under IRB approval from our University's Ethics Committee, with particular attention to the following ethical considerations operate under strict ethical guidelines with particular attention to participant comfort, data privacy, and the right to withdraw. Regular assessment of participant well-being and system performance is going to allow for responsive adjustments to the interaction protocol.

By conducting this study in community centers rather than homes, we will ensure standardized technical conditions while maintaining a familiar and comfortable environment for participants. The findings will guide technical refinements to our motion capture system and implementation strategies for broader deployment in elderly care settings.

\section{Limitation and Future work}
Despite the planned user study, we acknowledge several limitations in our current implementation and identify numerous opportunities for future improvement.

\textbf{Posture Recognition:} One of the primary challenges in posture recognition lies in resolving ambiguities during the dimensional elevation process from 2D images to 3D, as well as addressing issues like occluded key points and invisible body parts. Our AI system preprocesses camera-captured images and performs spatial calculations to extract precise 3D spatial data. To achieve this, we have trained our model on a dataset containing over ten thousand hours of video data, utilizing convolutional neural networks and deep learning algorithms. While this system effectively reconstructs and recognizes human motion by incorporating principles of dynamics and biomechanics, further refinement is needed to enhance accuracy in complex real-world environments, such as crowded or poorly lit settings.

\textbf{Behavior Recognition:} The goal of behavior recognition is to enable machines to accurately interpret human actions. This requires categorizing common behaviors and defining emotional actions with distinct and meaningful labels. For our current system, we trained datasets using one hundred thousand labeled images for each predefined behavior. To deepen our understanding of human behavior, we also incorporated multiple multimodal datasets that analyze combinations of key behavioral actions. However, our current approach is limited by predefined categories, and future work could explore more dynamic and flexible classification methods to better capture nuanced and spontaneous behaviors.

\textbf{Multimodal Fusion:} Multimodal fusion is critical for enabling AI to respond effectively to a variety of user actions. Our system extracts meaningful features by leveraging the complementary semantic information between visual and auditory inputs, integrating these features during an early fusion stage. While the behavior recognition module identifies emotional labels based on visual data, the audio input model—deployed locally as a miniature model—analyzes complex contexts and semantics to determine the user’s current emotional state. Despite its functionality, the system could benefit from more advanced fusion techniques to handle a broader range of modalities and enhance its robustness in noisy or unpredictable environments.

Moving forward, we aim to address these limitations by improving:

\begin{itemize}

\item \textbf{Posture Recognition:} Developing more sophisticated algorithms to handle occlusions and improving dimensional elevation accuracy, particularly in challenging scenarios.
\item \textbf{Expanding Behavior Recognition:} Introducing adaptive learning mechanisms to dynamically capture and classify a wider range of behaviors and emotional expressions.
\item \textbf{Advancing Multimodal Fusion:} Exploring state-of-the-art fusion techniques, such as transformer-based models, to better integrate multimodal data and enhance the system’s contextual understanding.
\item \textbf{Real-World Validation:} Conducting extensive user studies to validate system performance and usability in diverse real-world settings, allowing us to refine our approach based on feedback and practical challenges.
\end{itemize}

By addressing these areas, we aim to create a more robust, intuitive, and empathetic virtual companion system that better supports the needs of elderly users.

\section{Conclusion}

In this research, we introduced CyanKitten, an interactive virtual companion system designed specifically for elderly users. The system integrates advanced posture recognition, behavior recognition, and multimodal interaction capabilities. Our work demonstrates the technical feasibility of leveraging artificial intelligence to foster meaningful social interactions for older adults through virtual companionship. By addressing key technical challenges in elder-focused companion systems, we developed a solution that prioritizes usability and engagement.

The posture recognition system is tailored to capture and interpret the unique movement patterns of older adults, while the behavior recognition component is designed to identify and respond to specific interactive gestures. Additionally, the multimodal integration layer enhances the interaction experience by combining real-time visual and auditory inputs, enabling natural and intuitive exchanges between users and their virtual companions.

To evaluate the effectiveness and impact of CyanKitten, we are planning comprehensive user studies with elderly participants. These studies will assess the system's usability, engagement levels, and its potential to improve social and emotional well-being. The evaluation will include both quantitative metrics (e.g., system performance) and qualitative assessments (e.g., user experience and satisfaction). Through these studies, we aim to gain valuable insights into how elderly users interact with virtual companions in real-world settings.

We also identified future development paths to advance this technology further. These include enhancing system robustness, expanding interaction capabilities, and improving deployment feasibility. Insights from the planned user studies will guide these refinements, ensuring that future iterations align with the needs and preferences of elderly users.

Our research establishes a technical foundation for AI-powered companion systems specifically designed for older adults. As we move forward with user evaluations and system enhancements, we believe this work will contribute significantly to the growing field of elderly-focused interactive technologies. The insights gained will inform the development of more advanced, empathetic, and accessible virtual companions capable of addressing the social and emotional needs of aging populations.

\bibliographystyle{ACM-Reference-Format}
\bibliography{sample-base}

\appendix

\end{document}